\documentclass[aps,prb,twocolumn,groupedaddress,showpacs]{revtex4}

\usepackage{graphicx}
\usepackage{dcolumn}
\usepackage{bm}

\begin{document}

\title{Orbital-controlled magnetic transition between gapful and gapless
phases \\ in the Haldane system with $t_{\rm 2g}$-orbital degeneracy}

\author{Hiroaki Onishi and Takashi Hotta}
\affiliation{Advanced Science Research Center,
Japan Atomic Energy Research Institute,
Tokai, Ibaraki 319-1195, Japan}

\date{August 23, 2004}

\begin{abstract}
In order to clarify a key role of orbital degree of freedom
in the spin $S$=1 Haldane system, we investigate ground-state properties
of the $t_{\rm 2g}$-orbital degenerate Hubbard model
on the linear chain by using numerical techniques.
Increasing the Hund's rule coupling in multi-orbital systems,
in general, there occurs a transition from an antiferromagnetic
to a ferromagnetic phase.
We find that the antiferromagnetic phase is described as the Haldane
system with spin gap, while in the ferromagnetic phase, there exists
the gapless excitation with respect to orbital degree of freedom.
Possible relevance of the present results to actual systems
is also discussed.
\end{abstract}


\maketitle


Since the Haldane conjecture,\cite{Haldane}
quasi-one-dimensional Heisenberg antiferromagnets with spin $S$$>$1/2
have attracted considerable attention
in the research field of condensed matter physics.
At first, it was unacceptable that the integer-spin chain exhibits
an energy gap in the spin excitation and the spin correlation
decays exponentially in contrast to the half-odd-integer-spin chain.
However, the prediction has been confirmed to be correct
after examined by theoretical,\cite{AKLT,Kennedy-Tasaki,Oshikawa}
numerical,\cite{QMC-S1,ED-S1,DMRG-S1}
and experimental investigations.\cite{NENP-1,NENP-2,NENP-3}
Especially, the valence-bond-solid (VBS) picture has clarified
the microscopic mechanism of the gapped ground state,\cite{AKLT}
and the Haldane gap has been actually observed
for spin $S$=1 compounds such as
Ni(C$_{2}$HgN)$_{2}$NO$_{2}$(ClO$_{4}$) abbreviated NENP.\cite{NENP-1}
It has been also found that when a magnetic field is applied to NENP,
finite magnetic moment begins to grow
at a critical field corresponding to the Haldane gap.\cite{NENP-2}

Now the existence of the Haldane gap is
confirmed in the spin $S$=1 system, but in LiVGe$_2$O$_6$,
a quasi-one-dimensional spin $S$=1 system,
antiferromagnetic (AFM) transition occurs at a N\'eel temperature
$T_{\rm N}$=22K and the expected Haldane gap is absent
or strongly suppressed.\cite{Millet}
To understand the suppression contrary to the Haldane
conjecture, a scenario has been proposed based on
biquadratic interaction of neighboring spins.
In fact, the susceptibility has been reproduced
by the bilinear-biquadratic $S$=1 Heisenberg chain,
in combination with the effect of impurity contribution.\cite{Lou}
The biquadratic interaction has been derived by
the fourth-order perturbation in terms of the electron
hopping,\cite{Milla} assuming a significant value
of the level splitting in $t_{\rm 2g}$ orbitals of V$^{3+}$ ion.
However, the splitting has been found to be much smaller than
previously considered,\cite{Vonlanthen}
indicating that orbital fluctuations may play a crucial
role in LiVGe$_2$O$_6$.

Here let us reconsider the Haldane system from an electronic viewpoint.
As recognized in the VBS picture, spin $S$=1 is decomposed
into a couple of $S$=1/2 electrons.
In actual systems, the spin $S$=1 state is stabilized
by the Hund's rule coupling among electrons in different orbitals.
For Ni$^{2+}$ ion in NENP, two electrons in $e_{\rm g}$ orbitals
form $S$=1 and orbital degree of freedom is inactive.
On the other hand, V$^{3+}$ ion
contains two electrons in three $t_{\rm 2g}$ orbitals.
When the level splitting among $t_{\rm 2g}$ orbitals
is small, as actually observed in LiVGe$_2$O$_6$,
orbital degree of freedom remains active
after forming $S$=1, in contrast to Ni$^{2+}$ ion.
Significance of $t_{\rm 2g}$-orbital degree of
freedom has been also emphasized to understand the peculiar
magnetic behavior of cubic vanadate YVO$_3$.\cite{Sirker}
Similar situation may occur in one-dimensional ruthenium compounds,
since the low-spin state of Ru$^{4+}$ ion with $4d^{4}$
configuration includes two holes.
In the Haldane system, we have learned that distinctive features
originate from strong quantum effects of the spin.
We believe that it is fascinating to consider further
orbital degree of freedom in the Haldane system.
In particular, it is important to clarify how orbital degeneracy
affects magnetic properties such as the gapful nature
of the spin-only system.

In this paper, we investigate ground-state properties of
the $t_{\rm 2g}$-orbital degenerate Hubbard model on the linear chain
by numerical techniques such as the Lanczos diagonalization
and the density matrix renormalization group (DMRG) method.\cite{White}
When the Hund's rule coupling is small, an AFM phase with ferro-orbital
(FO) ordering appears.
This phase is well described by the spin $S$=1 AFM Heisenberg model
and thus, the Haldane gap exists.
On the other hand, increasing the Hund's rule coupling,
there occurs a ferromagnetic (FM) phase with antiferro-orbital (AFO)
correlation, in which the system is described by
the pseudo-spin $T$=1/2 AFO Heisenberg model
with the gapless orbital excitation.
Namely, the low-energy physics drastically changes
between gapful and gapless phases as a result of orbital ordering.


Let us consider a chain of ions including four electrons per site, where
the local spin $S$=1 state is expected to form in $t_{\rm 2g}$ orbitals.
Here we envisage Ru$^{4+}$ chain, but due to the electron-hole symmetry,
the results are also applicable to V$^{3+}$ chain with two electrons
in $t_{\rm 2g}$ orbitals per site.
Note that the direction of the chain is set to be the $x$-axis,
but the result does not depend on the chain direction
due to the cubic symmetry.
The orbital degenerate Hubbard model is given by
\begin{eqnarray}
 \label{eq-H}
 H &=& \sum_{i,\gamma,\gamma',\sigma}
 t_{\gamma\gamma'} d_{i\gamma\sigma}^{\dag} d_{i+1\gamma'\sigma}
 \nonumber\\
 &+& U \sum_{i,\gamma} \rho_{i\gamma\uparrow} \rho_{i\gamma\downarrow}
 +U'/2 \sum_{i,\sigma,\sigma',\gamma \ne \gamma'} 
 \rho_{i\gamma\sigma} \rho_{i\gamma'\sigma'} \nonumber \\
 &+& J/2 \sum_{i,\sigma,\sigma',\gamma \ne \gamma'} 
 d_{i\gamma\sigma}^{\dag} d_{i\gamma'\sigma'}^{\dag}
 d_{i\gamma\sigma'} d_{i\gamma'\sigma} \nonumber \\
 &+& J'/2 \sum_{i,\sigma \ne \sigma',\gamma \ne \gamma'} 
 d_{i\gamma\sigma}^{\dag} d_{i\gamma\sigma'}^{\dag}
 d_{i\gamma'\sigma'} d_{i\gamma'\sigma},
\end{eqnarray}
where $d_{i\gamma\sigma}$ is an annihilation operator
for an electron with spin $\sigma$ in the orbital $\gamma$
(=$xy$, $yz$, $zx$) at site $i$,
$t_{\gamma\gamma'}$ is the nearest-neighbor hopping between
adjacent $\gamma$ and $\gamma'$ orbitals, and
$\rho_{i\gamma\sigma}$=$d_{i\gamma\sigma}^{\dag}d_{i\gamma\sigma}$.
Note that hopping amplitudes are given as
$t_{xy,xy}$=$t_{zx,zx}$=$t$ and zero for other combinations of orbitals.
Since we consider the chain along the $x$-axis,
there is no hopping for $yz$ orbitals.
Hereafter, $t$ is taken as the energy unit.
In the interaction terms, $U$ is the intra-orbital Coulomb interaction,
$U'$ the inter-orbital Coulomb interaction,
$J$ the inter-orbital exchange interaction (the Hund's rule coupling),
and $J'$ is the pair-hopping amplitude between different orbitals.
Note the relation $U$=$U'$+$J$+$J'$ due to the rotation symmetry
in the orbital space and $J'$=$J$ is also assumed
due to the reality of the wavefunction.

\begin{figure}[t]
\includegraphics[width=0.95\linewidth]{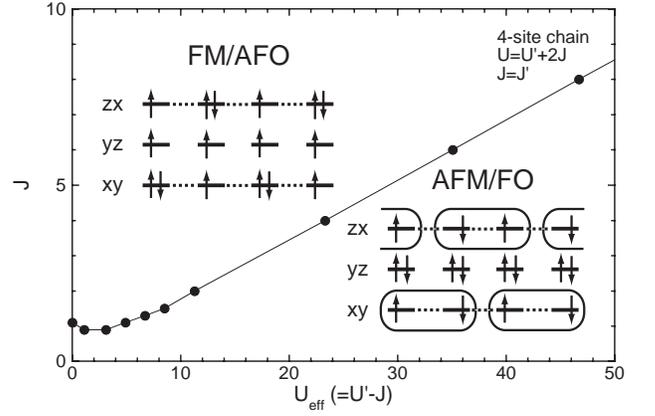}
\caption{\label{fig-4site}
Ground-state phase diagram for the 4-site chain.
Insets indicate schematic views for electron configurations
in the AFM and FM phases.
Oval enclosures represent the spin singlet pair
for the VBS picture.}
\end{figure}

First we show the results of the Lanczos diagonalization for
the 4-site chain with the anti-periodic boundary condition.
Although the system size is very small and it is necessary to
pay due attention to a relation with actual materials,
there are advantages that we can quickly accumulate lots of results
and easily gain insight into ground-state properties.
In Fig.~\ref{fig-4site}, the ground-state phase diagram is shown
in the ($U_{\rm eff}$, $J$) plane with $U_{\rm eff}$=$U'$$-$$J$.
The phase boundary is determined by comparing the energies
for the phases of $S_{\rm tot}^{z}$=0$\sim$4,
where $S_{\rm tot}^{z}$ is the $z$-component of total spin.
Here two types of the ground state are observed.
In the region of small $J$, the lowest-energy state is characterized
by $S_{\rm tot}^{z}$=0, indicating the spin singlet state
regarded as the one-dimensional AFM phase.
On the other hand, the maximally spin-polarized FM phase appears
in the region of large $J$, where the energies for the phases with
$S_{\rm tot}^{z}$=0$\sim$4 are degenerate.

In order to obtain an intuitive explanation for the appearance of
both phases, we show the electron configuration in $t_{\rm 2g}$
orbitals for the AFM and FM phases in the insets of Fig.~1.
In the AFM phase, since there is no hopping between adjacent $yz$ orbitals
along the $x$-axis, $xy$ and $zx$ orbitals should be singly occupied
to gain the kinetic energy by electron hopping, while $yz$ orbital is
doubly occupied and localized.
The Hund's rule coupling stabilizes the local spin $S$=1 state
consisting of two electrons with parallel spin
in $xy$ and $zx$ orbitals,
which are antiferromagnetically coupled between nearest neighbor sites.
Note that the electron configuration in the AFM phase is well understood
from the VBS picture:
Due to the hopping connection, it is natural that a spin singlet pair
should be formed on adjacent $xy$ and $zx$ orbitals.
Considering that the spin singlet pair is alternately assigned to
$xy$ or $zx$ orbital, the uniform VBS state is constructed.
On the other hand, in the FM phase, three up-spin electrons
occupy different orbitals due to the Pauli principle.
One down-spin electron can be accommodated in any of three orbitals
in principle, but in order to gain the kinetic energy,
down-spin electrons occupy $xy$ or $zx$ orbital alternately.
By introducing $T$=1/2 pseudo-spin operators representing
$xy$ and $zx$ orbitals,
the AFO structure is expressed as an alternating correlation
of the pseudo-spin.

From the above discussion on the electron configuration,
it is useful to consider the effective model
in the strong-coupling region to clarify ground-state properties.
For the AFM phase, the effective model is given by
\begin{equation}
 \label{eq-Heff-AFM}
 H_{\rm AFM} = J_{\rm AFM} \sum_{i} 
 ({\bf S}_{i} \cdot {\bf S}_{i+1}-1),
\end{equation}
where ${\bf S}_i$ indicates the $S$=1 operator and the
AFM coupling $J_{\rm AFM}$ is expressed as
$J_{\rm AFM}$=$2t^2/(U+J)$.
On the other hand, the effective model in the FM phase
is given by the $T$=1/2 AFO Heisenberg model
\begin{equation}
 \label{eq-Heff-FM}
 H_{\rm FM}=2J_{\rm FM} \sum_{i} 
 ({\bf T}_{i}\cdot{\bf T}_{i+1}-1/4),
\end{equation}
with $J_{\rm FM}$=$2t^2/U_{\rm eff}$.
We define pseudo-spin operator as
${\bf T}_{i}$=$(1/2)\sum_{\gamma,\gamma',\sigma}
d_{i\gamma\sigma}^{\dag}\bm{\sigma}_{\gamma\gamma'}d_{i\gamma'\sigma}$,
where $\bm{\sigma}$=$(\sigma_x, \sigma_y, \sigma_z)$ are Pauli matrices
and the summation for $\gamma$ and $\gamma'$ is
taken for $xy$ and $zx$ orbitals.
Here we arrive at the following interesting conclusion:
The AFM phase of the $t_{\rm 2g}$-orbital degenerate
Hubbard model is the $gapped$ Haldane phase,
well described by the spin $S$=1 AFM Heisenberg model $H_{\rm AFM}$.
On the other hand, the FM phase is described by
the pseudo-spin $T$=1/2 AFO Heisenberg model $H_{\rm FM}$
with the $gapless$ excitation.


We believe that the characteristics grasped within the small
4-site chain are physically meaningful, but it is highly required
to confirm them without finite size effects.
In order to consider numerically the problem in the thermodynamic
limit, we employ the infinite-system DMRG method
with open boundary condition.\cite{White}
Note that one site consists of three $t_{\rm 2g}$ orbitals
and the number of bases is 64 for one site,
indicating that the matrix size becomes very large.
In the present calculation, the number of states kept for each
block $m$ is up to $m$=160 and the truncation error is estimated
as $10^{-5}$ at most.

\begin{figure}[t]
\includegraphics[width=0.95\linewidth]{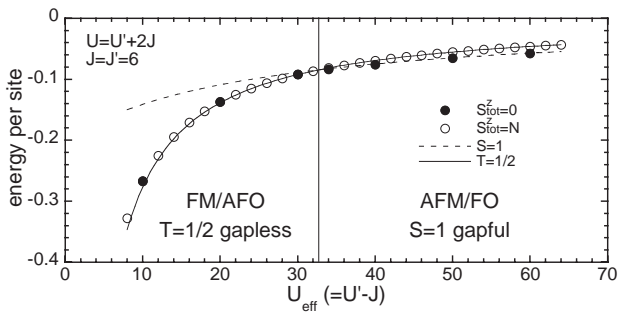}
\caption{\label{fig-gse}
Ground-state energy per site as a function of $U_{\rm eff}$ for
$J$=6 in the thermodynamic limit. Note that the origin of the energy
is shifted (see the maintext).
Dashed and solid curves denote exact ground-state energies for
the $S$=1 AFM and $T$=1/2 AFO Heisenberg chains, respectively.}
\end{figure}

In Fig.~\ref{fig-gse}, the ground-state energy is plotted for
$S_{\rm tot}^z$=0 and $N$ (the number of sites, i.e.,
the maximally polarized state).
Here we fix $J$=6 to consider the strong-coupling region.
Note that the energy is shifted by the constant of on-site Coulomb
interactions $5U'+U-3J$ for comparison with the effective model.
In the AFM phase, the lowest-energy state is characterized by
$S_{\rm tot}^z$=0 (solid circle),
which is in agreement with the exact energy of $H_{\rm AFM}$
(dashed curve).\cite{note1}
On the other hand, the energies for $S_{\rm tot}^z$=0 and $N$
are degenerated in the FM phase.
The DMRG results (open circle) are compared with
the exact energy of $H_{\rm FM}$ (solid curve).\cite{note2}
The agreement is fairly well, except for a little discrepancy
for small $U_{\rm eff}$,
since the strong-coupling limit is appropriate for large $U_{\rm eff}$.
Therefore, we conclude that the ground-state energy is well reproduced
by the effective model in the strong-coupling region.
Note that the phase transition takes place
at a critical point $U_{\rm c}$$\simeq$33, which is similar to
the value of the 4-site chain $U_{\rm c}(N$=$4)$$\simeq$35.

\begin{figure}[t]
\includegraphics[width=0.95\linewidth]{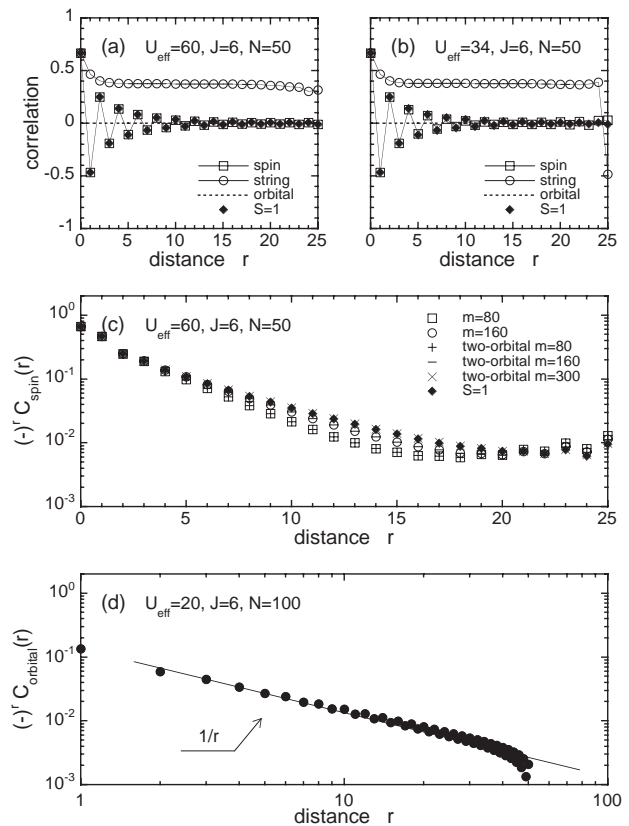}
\caption{\label{fig-corr}
(a) Spin, string, and orbital correlation functions measured
from the center of the chain for the AFM state.
Here $r$ denotes the distance from the center position.
Solid diamonds denote the spin correlation function of $H_{\rm AFM}$.
(b) Correlation functions for $U_{\rm eff}$=34.
(c) The linear-log plot of the spin correlation function
in Fig.~\ref{fig-corr}(a) and the two-orbital model.
(d) The log-log plot of the orbital correlation function
for the FM state.}
\end{figure}


In order to verify that ground-state properties are described by
the effective model from the microscopic viewpoint,
it is quite important to investigate correlation functions.
For the AFM state, we evaluate the spin correlation function,
$C_{\rm spin}(j,k)$=$\langle S_{j}^{z}S_{k}^{z} \rangle$,
the string correlation function,\cite{Nijs-Rommelse}
$C_{\rm string}(j,k)$=$\langle
S_{j}^{z}\exp[i\pi\sum_{l=j}^{k-1}S_{l}^{z}]S_{k}^{z} \rangle$,
and the orbital correlation function,
$C_{\rm orbital}(j,k)$=$\langle T_{j}^{z}T_{k}^{z} \rangle$.
Here $\langle \cdots \rangle$ indicates the average using
the ground-state wavefunction.
In Fig.~\ref{fig-corr}(a),
we show correlation functions for $U_{\rm eff}$=60.
We also depict $C_{\rm spin}$ of $H_{\rm AFM}$
obtained by the quantum Monte Carlo (QMC) method
with the loop algorithm.\cite{note-qmc}
We observe that $C_{\rm spin}$ rapidly decays
in accordance with the QMC results
and the string long-range order exists,
indicating that the characteristics of $H_{\rm AFM}$ are reproduced
and the system has the {\it gapful} nature with respect to the spin.
Note that the orbital has no effect on the low-lying excitation
due to the localization of $yz$ orbital,
as shown by dashed line in Fig.~\ref{fig-corr}(a).
For $U_{\rm eff}$=34 (near the phase boundary),
the behavior of correlation functions in the bulk
is also described by $H_{\rm AFM}$,
as shown in Fig.~\ref{fig-corr}(b).
Note that anomaly due to open boundary condition appears
just at the chain edge,
since an orbital singlet dimer state with FM spin structure
is stabilized to gain energy locally at the chain edge.

In Fig.~\ref{fig-corr}(c),
we show $C_{\rm spin}$ in the linear-log scale,
indicating evidence of the exponential decay, although
there is slight deviation from that of the QMC result
for $H_{\rm AFM}$.
In the DMRG calculation for $H$, the truncation error is about
10$^{-7}$ for $m$=160 and the energy is evaluated in this accuracy.
However, in order to evaluate $C_{\rm spin}$
in the same accuracy, it is necessary to increase $m$ further and
check the convergence with respect to $m$.
Unfortunately, it is quite difficult to perform such tasks
due to the large matrix size.
Instead, here we introduce a two-orbital Hubbard model composed
of $xy$ and $zx$ orbitals, in order to reduce the matrix dimension
effectively.
When we compare the results of the effective two-orbital model
and the original Hamiltonian for $m$=80 and 160,
good agreement is obtained
as shown in Fig.~\ref{fig-corr}(c),
indicating that the two-orbital model well reproduces
ground-state properties of $H$ in the AFM region.
Further increasing $m$ up to $m$=300 in the two-orbital model
to improve the precision of the finite-system DMRG method,
we confirm that $C_{\rm spin}$ of the two-orbital model
converges to that of $H_{\rm AFM}$.
Then, it is highly believed that the same convergence should be
observed for the original $t_{\rm 2g}$-orbital model.

For the FM state, we evaluate $C_{\rm orbital}$.
Note that $C_{\rm spin}$ is constant
depending on $S_{\rm tot}^z$ in the maximally spin-polarized FM phase.
In Fig.~\ref{fig-corr}(d), we observe
the power-law decay of $C_{\rm orbital}$,
which is characteristic of the half-odd-integer-spin chain.
Namely, the system is described by $H_{\rm FM}$
and quantum fluctuation with respect to the orbital
gives the {\it gapless} excitation.
Further, the spin-wave excitation presents due to FM ordering
and thus, both spin- and orbital-wave excitations are expected
to exist.


Now we discuss relevance of our results to actual systems.
We have performed the DMRG calculations for large value of $J$=6,
to discuss the behavior in the strong-coupling region.
However, we have confirmed that for the small 4-site chain,
spin and orbital structures show characteristic behavior in each phase
even in the weak-coupling region with small $U_{\rm eff}$ and $J$.
In addition, as mentioned above, the phase boundary for $J$=6 is almost
the same as that obtained in the 4-site chain.
Namely, it is expected that the phase diagram in Fig.~\ref{fig-4site}
does not change so much even in the thermodynamic limit,
although the AFM region may somewhat extend.
If we assume $t$=0.4eV, $U$=7eV, and $J$=1eV as typical values
for transition metal oxides,\cite{Dagotto-review}
we obtain $U_{\rm eff}/t$=10 and $J/t$=2, where the system is
expected to be near the phase boundary.
Thus, the transition from gapful AFM to gapless FM
phases may be controlled by chemical pressure.

In addition to the control of the Coulomb interactions or the bandwidth,
it is possible that the magnetic field induces ferromagnetism
in the Haldane system.\cite{comment}
The system is in the singlet ground state below a critical field
due to the Haldane gap, indicating that magnetic moment is suppressed.
At the critical field, finite value of magnetic moment begins to grow.
Note that in actual systems, spin-canted AFM phase occurs.
Increasing further the magnetic field, the saturated FM state should occur,
when we apply a high magnetic field corresponding to
4$J_{\rm AFM}$.\cite{note3}
For instance, this value may be estimated as 200 Teslas for LiVGe$_2$O$_6$,
but smaller in other Haldane systems.
The gapless excitation with respect to the orbital
is expected to be observed in a similar way that
the orbital-wave excitation has been revealed
by Raman scattering measurements in LaMnO$_{3}$.\cite{orbiton}


In summary, ground-state properties of the $t_{\rm 2g}$-orbital degenerate
Hubbard model have been investigated in order to clarify effects of orbital
degeneracy in the spin $S$=1 Haldane system.
In the AFM phase, the spin $S$=1 AFM Heisenberg model is reproduced,
while in the FM phase, the system is described by
the pseudo-spin $T$=1/2 AFO Heisenberg model.
Thus, we conclude that the low-energy physics is drastically changed
due to the interplay of spin and orbital degrees of freedom.
In particular, the gapless orbital excitation exists in the FM state.
It may be interesting to observe such orbital excitation
in the Haldane system under the high magnetic field.


The authors thank K. Kakurai and K. Yoshimura for discussions
and comments.
One of the authors (T.H.) is supported by a Grant-in-Aid from
the Ministry of Education, Culture, Sports, Science and Technology
of Japan.



\begin{thebibliography}{99}
\bibitem{comment}
Note that this is $not$ the direct transition between
the AFM and FM phases. Here we emphasize that
$t_{\rm 2g}$-orbital degeneracy could remain effective
in the FM phase of the Haldane system,
leading to gapless excitation.
\bibitem{note3}
The saturation field is given by the energy difference between
the fully polarized states with the total spin
$S_{\rm tot}$=$N$ and $N$$-$1,
i.e., $[E(N)-E(N-1)]$/$J_{\rm AFM}$=$N-(N-4)$=4.
\end{thebibliography}
\end{document}